\documentstyle[twocolumn,aps,prl,epsf,floats]{revtex}
\def\lsi{\raise0.3ex
\hbox{$<$\kern-0.75em\raise-1.1ex\hbox{$\sim$}}}
\def\gsi{\raise0.3ex
\hbox{$>$\kern-0.75em\raise-1.1ex\hbox{$\sim$}}}

\def\vf{\phi}

\begin{document}
\twocolumn[\hsize\textwidth\columnwidth\hsize\csname
@twocolumnfalse\endcsname

\title{On the variation of the gauge couplings during inflation}
\author{Massimo Giovannini}
\address{{\it Institute of Theoretical Physics, University of Lausanne, 
CH-1015 Lausanne, Switzerland}}
\maketitle

\begin{abstract}
\noindent
It is shown that the evolution of the (Abelian) 
gauge coupling during an inflationary phase of de 
Sitter type drives the growth of the two-point function
of the magnetic inhomogeneities.  
After examining the  constraints on the variation of the gauge coupling 
arising in a standard model of inflationary 
and post-inflationary evolution,
magnetohydrodynamical 
equations are generalized to the case of time evolving gauge coupling.
It is argued that large scale magnetic fields can be copiously 
generated. Other possible implications of the model are outlined.

\end{abstract}

\vskip1.5pc]

\section{Introduction} 

Prior to the formation of the light 
elements (taking place
at a temperature of roughly $0.1$ MeV)
the gauge coupling could have  been dynamical \cite{dir,bar}.
Two examples in this direction 
are models involving extra-dimensions \cite{a} and string motivated 
scenarios \cite{b}. 

Suppose that a minimally coupled (massive) scalar field $\phi$ evolves in a 
conformally flat metric of Friedmann-Robertson-Walker type:
\begin{equation}
ds^2 = G_{\mu\nu} dx^{\mu} dx^{\nu}=
 a^2(\eta) [ d\eta^2 - d\vec{x}^2],
\label{metric}
\end{equation} 
where $a(\eta)$ is the scale factor, $\eta$ the conformal time 
coordinate [related to cosmic time $t$ 
as $ a(\eta) d\eta = dt$] and $G_{\mu\nu}$ the space-time metric. 
The field $\phi$ {\em is not the inflaton} but it evolves 
during different cosmological epochs parametrised by a 
different form of $a(\eta)$. Typically the Universe 
evolves from an inflationary phase of de Sitter (or quasi-de Sitter)
 type towards
a radiation dominated phase which is finally replaced
by a matter dominated epoch.

The evolution equation of $\phi$ in the background given by Eq. (\ref{metric}) 
can be written as 
\begin{equation}
\phi'' + 2 {\cal H} \phi' + m^2 a^2 \phi=0,\,\,\,{\cal H} = \frac{a'}{a},
\label{ph}
\end{equation}
where the prime denotes derivation with respect to the conformal time 
coordinate and 
${\cal H}$ is the Hubble factor in conformal time related to 
the Hubble parameter in cosmic time $H = \dot{a}/a$ as $ H a ={\cal H}$
(the dot denotes derivative with respect to cosmic time).

If $\phi$ evolves during an inflationary phase 
of de Sitter type the scale factor will be  
\begin{equation}
a(\eta) = \bigl( -\frac{\eta}{\eta_1})^{-1}, \,\,\,\eta < - \eta_1,
\label{inf}
\end{equation}
where $-\eta_1$ marks the end of the inflationary phase. 
If $m^2 a^2 \ll {\cal H}$ (i.e. $m \ll H$) during inflation, according 
to Eq. (\ref{ph}),  $\phi$ relaxes as 
$\phi \sim (-\eta/\eta_1)^3$ for $\eta < - \eta_1$.

Suppose, as an example, 
 that $\phi$ is coupled to an (Abelian) gauge field  
\begin{equation}
S \sim \int d^4 x \sqrt{-G} \phi^2 F_{\mu\nu}F^{\mu\nu}.
\label{ac}
\end{equation}
The normal modes of the hypermagnetic field fluctuations 
$B_{i}(\vec{x}, \eta)$ 
are $b_{i}(\vec{X},\eta) = \phi(\eta) B_{i}(\vec{x},\eta)$ and their 
correlation function during the de Sitter phase can then be written as
\begin{equation}
{\cal G}_{ij} (\vec{r},\eta) = \int \frac{d^3 k}{(2\pi)^3} P_{i j}(k)
 b(k,\eta) b^{\ast}(k\eta) 
e^{i \vec{k}\cdot \vec{r}},
\label{two}
\end{equation}
where 
\begin{equation}
P_{i j} = \biggl( \delta_{i j} - \frac{k_{i} k_{j}}{k^2} \biggr).
\end{equation}
The normal modes  $b(k,\eta)$ will evolve as 
\begin{equation}
b'' +\bigl[k^2 - \frac{\phi''}{\phi}\bigr] b =0.
\end{equation}
Using now the fact that $\phi \sim \eta^3$ the correlation function, 
during the de Sitter phase grows as
\begin{equation}
{\cal G}_{i j} (\vec{r},\eta) \sim (-\eta)^{-4},
\end{equation}
for $\eta \rightarrow 0^{-}$ (i.e. $t\rightarrow\infty$). 
Thus, gauge field fluctuations grow during the Sitter stage. Furtheremore,
from Eq. (\ref{ac}) the magnetic energy density $\rho_{\rm B}(r,\eta)$
[related to the 
trace of ${\cal G}_{ij}(\vec{r},\eta)$]  also
increases for $\eta\rightarrow 0^{-}$. Consequently, since
the magnetic energy density can be amplified during a de Sitter-like 
stage of expansion, large scale gauge fluctuations 
pushed outside of the horizon can generate the galactic magnetic field. 

\section{Evolution of the gauge coupling}

The only gauge coupling free to evolve, in the present discussion, 
 is the one associated 
with the hypercharge field leading, after symmetry breaking, to the time 
variation of the electron charge. In 
 a relativistic plasma the conductivity goes,  
approximately, as $T/\alpha_{\rm em}$ where $\alpha_{\rm em}$ 
is the fine structure constant \cite{k}. If $\alpha_{\rm em}$ depends on time 
also the well known magnetohydrodynamical equations (MHD) \cite{k} 
will have to be generalized,  leading, ultimately, 
to different mechanism for the relaxation of the 
magnetic fields. 

If the evolution of the Abelian coupling is parametrised through the 
minimally coupled scalar field $\phi$, the possible 
constraints pertaining to the evolution of $\phi$ are translated 
into constraints on the evolution of the gauge coupling.
Massless scalars cannot exist in the Universe: they lead 
to long range forces whose effect should appear in 
sub-millimiter tests of Newton's law. 
Consequently, the scalar 
mass should be, at least, larger than $10^{-4}$ eV otherwise 
it would be already excluded \cite{mm}. 
Massive scalars are severely 
constrained from cosmology \cite{r,l}. When 
the scalar mass is comparable with the Hubble 
rate (i.e. $m \sim H$) the 
field starts oscillating coherently 
with Planckian amplitude and $\phi$
decays too late big-bang nucleosynthesis (BBN) 
can be spoiled \cite{bar}.

Initial conditions for the evolution 
of $\phi$ are given during a de Sitter 
stage of expansion. Thus, the homogeneous evolution of 
$\phi$ can be written as 
\begin{equation}
\phi_{\rm i}(\eta) \sim \phi_1 - \phi_2 \bigl( \frac{\eta}{\eta_1}
 \bigr)^3,\,\,\,\eta < -\eta_1,
\end{equation}
where $\phi_1$ is the asymptotic value of $\phi$ which may or may 
not coincide with the minimum of $V(\phi)$; $\phi_2$ is 
also an integration constant. Without fine-tuning 
$\phi_1$ and $\phi_2$ both coincide with $M_{P}$.
During the inflationary phase 
$m\ll {\cal H}_1/a_1 = H_1$ where, $H_1 < 10^{-6} \,M_{P}$ is the 
curvature scale at the end of inflation.

After $\eta_1$ the Universe enters a phase of radiation
dominated evolution  (possibly preceded by a 
reheating phase) where the curvature scale 
decreases. When $H_{\rm m} \sim m$  
the scalar field starts oscillating coherently 
with  amplitude $\phi_1$. 

During reheating the scale factor evolves 
as $a(\eta) \sim \eta^{\alpha}$ so that, in this 
phase, $\phi$ relaxes as 
\begin{equation}
\phi_{\rm rh} 
\sim \phi_1 + \bigl(\frac{\eta}{\eta_1})^{ 1 - 2\alpha},\,\,\eta_1 < 
\eta <\eta_r,
\label{rh}
\end{equation}
where $\eta_{\rm r}$ marks the beginning of the radiation 
dominated phase occurring at a scale $H_{\rm r}>m$. 
 In the case of matter-dominated equation of 
state during reheating $\alpha \sim 2 $.

For $\eta>\eta_r$, the evolution of the field 
$\phi$ can be exactly solved 
(in cosmic time) in terms of Bessel functions
\begin{equation}
\phi(t) \sim a^{-3/2}(t) 
\sqrt{m t} \bigl[ A Y_{1/4}(m t) + B J_{1/4}(m t)\bigr],
\label{ra}
\end{equation}
where $A$ and $B$ are two integration constants.
From Eq. (\ref{ra}), $\phi \sim {\rm constant} +\eta^{-1} $
for $H> m$, and it oscillates for $H < m$. 
When $H <m$ the coherent oscillations of $\phi$ start and 
their energy density  decreases as
$a^{-3}$.
The curvature scale $H_{\rm c}$ marks the time 
at which the energy density stored in the 
 coherent oscillations equal 
the energy density of the radiation background,  namely
\begin{equation}
H_{\rm r}^2 M_{P}^2 \bigl(\frac{a_{\rm r}}{a_{\rm c}}\bigr)^4 \sim 
m^2 \phi_1^2 \bigl(\frac{a_{\rm m}}{a_{\rm c}}\bigr)^3,
\label{eq}
\end{equation}
where $\eta_{\rm m}$  
correspond to the times at which $H\sim m$. From Eq. (\ref{eq})
\begin{equation}
H_{\rm c} \sim \xi\, \varphi^4\,M_{P}.
\end{equation}
where $\varphi = \phi_1/M_{P}$ and $\xi = m/M_{P}$.
The phase of dominance of coherent oscillation ends with
the decay of $\phi$ at a scale dictated by the strength 
of gravitational interactions and by the mass $m$, namely
\begin{equation}
H_{\phi} \sim \xi^3 \,M_{P}.
\end{equation}
In order not to spoil the light elements abundances 
we have to require that $H_{\phi} > H_{\rm ns} $
implying that $ m > 10$ TeV.  

In order not wash-out the baryon asymmetry produced
at the electroweak time by overproduction of entropy \cite{ew1,ew2}
$H_{\phi} > H_{\rm ew}$ may be imposed. Since 
$H_{\rm ew} \sim \sqrt{N_{\rm eff}} T^2_{\rm ew} / M_{P}$ 
[where $N_{\rm eff}=106.75$ is the effective number of (spin)
 degrees of freedom at $T_{\rm ew} \sim 100$ GeV] 
 we obtain $ m > 10^{5} $ TeV.
Notice, incidentally, that the time variation of the gauge 
couplings during the electroweak epoch (possibly in the presence 
of a hypermagnetic field \cite{h})
 has not been analyzed and it may be relevant 
in order to produce inhomogeneities at the onset of BBN \cite{ks}.

The inhomogeneous modes of 
$\phi$ should also be taken into account since we have 
to check that  
further constraints are not introduced. In order 
to find how many quanta of the field $\phi$ are produced 
by passing from the inflationary phase to a radiation 
dominated phase let us look at the sudden approximation for 
the transition of $a(\eta)$ \cite{ms}. 
Consider the first order fluctuations of the field $\vf$
\begin{equation}
\vf(\vec{x},\eta) = \vf(\eta) + \delta \vf(\vec{x},\eta),
\end{equation}
whose evolution equation is, in Fourier space,
\begin{equation}
\psi'' + 2 {\cal H} \psi' + [ k^2 + m^2 a^2] \psi =0,
\label{fluc}
\end{equation}
where $\psi(k,\eta)$ is the Fourier component 
of $\delta\vf(\vec{x}, \eta)$. 

In the limit $ k\eta_1 \ll 1$ the mean number of 
quanta created by parametric amplification of vacuum 
fluctuations \cite{ms} is
\begin{equation}
\overline{n}(k) \simeq |c_{-}(k)|^2= q |k \eta_1|^{-2 \lambda} 
\biggl(\frac{m}{H_1}\biggr)^{-1/2} 
\end{equation}
where $q$ is a numerical coefficient of the 
order of $10^{-2}$. 
the energy density of the created (massive) quanta 
can be estimated from 
\begin{equation}
d\rho_{\psi} = \frac{d^3 \omega}{(2\pi)^3} m \overline{n}(k),
\end{equation}
where $ \omega = k/a$  is the physical momentum. 
In the case of a de Sitter phase ($\lambda = 3/2$) the typical 
energy density of the produced fluctuations 
is 
\begin{equation}
\rho_{\psi}(\eta) \simeq 
q \,m\, H_1^3\, 
\biggl( \frac{m}{H_1}\biggr)^{-1/2} \biggl(\frac{a_1}{a}\biggr)^3
\end{equation}
Also the massive fluctuations may become dominant and we have to make 
sure that they become dominant after $\vf$ already decayed. Define 
$H_{\ast}$ as the scale at which the massive fluctuations become 
dominant with  respect to the radiation background.
The scale $H_{\ast}$ can be determined 
by requiring that $\rho_{\psi}(\eta_{\ast}) \simeq \rho_{\gamma} (\eta_{\ast})$
implying that 
\begin{equation}
m \,H_1^3 \biggl(\frac{m}{H_1}\biggr)^{-1/2} 
\biggl(\frac{a_1}{a_{\ast}}\biggr)^3 
\simeq H_1^2 \, M_{P}^2 \biggl(\frac{a_1}{a_{\ast}}\biggr)^4,
\end{equation}
which translates into
\begin{equation}
H_{\ast}  \simeq q \, \xi\,\epsilon^4\, M_{P},
\end{equation}
where $\epsilon= H_1/M_{P}$.
In order to make sure that the non-relativistic modes 
will become dominant after $\vf$ already decayed 
we have to impose that $H_{\ast} < H_{\vf}$ which means that
$m > 10^{2}$ TeV for $H_1\sim 10^{-6}\, M_{P}$ and which is less 
restrictive than the other constraints previously derived in this paper.

\section{Magnetogenesis}
The full action describing the problem of the evolution 
of the gauge coupling  in this simplified scenario is 
\begin{eqnarray}
&&S= \int d^4 x \,\,\sqrt{-G} \biggl[ \frac{1}{2} G^{\mu\nu} \partial_{\mu} 
\vf \partial_{\nu} \vf 
- V(\vf) 
\nonumber\\
&&- \frac{1}{4} f(\phi) F_{\mu\nu} F^{\mu\nu} 
\biggr].
\label{action1}
\end{eqnarray}
Using Eq. (\ref{metric}) the equations of motion become
\begin{eqnarray}
&& \vf'' +  2 {\cal H} \vf ' +a^2 \frac{\partial V}{\partial \phi} 
= -\frac{1}{2 a^2}
\frac{\partial f}{\partial \phi} \biggl[ \vec{B}^2 - \vec{E}^2 \biggr] 
\label{s1}\\
&& \frac{\partial{\vec{B}}}{\partial\eta} = -\vec{\nabla}
\times \vec{{E}} 
,~~~~~{\vec{\nabla}}\cdot {\vec{E}}=0,
\label{s1b}\\
&&\frac{\partial}{\partial\eta}\biggl[f(\phi) \vec{E}\biggr] +
 \vec{J} =
f(\phi) {\vec{\nabla}}\times \vec{B}, 
\label{s2}\\
&& {\vec{\nabla}}\cdot{\vec{B}}=0,~~~~~~
\vec{J}=\sigma ({\vec{E}} + \vec{v}\times{\vec{B}})
\label{s4}
\end{eqnarray}
(${\vec{B}}=a^2 \vec{{\cal B }}$, $\vec{E}=a^2
\vec{{\cal E}}$; $\vec{J}=a^3 \vec{j}$; $\sigma=
\sigma_{c} a$;  $\vec{{\cal
B}}$, $\vec{{\cal E}}$,
$\vec{j}$, $\sigma_{c}$ are  the flat-space quantities whereas
$\vec{B}$, $\vec{E}$, $\vec{J}$, $\sigma$ are the
curved-space ones; $\vec{v}$ is the bulk velocity of the plasma). 

In Eqs. (\ref{s1})--(\ref{s4}) the effect of the conductivity 
has been included. The current density [present in Eq. (\ref{s1}) with a term 
$(\partial j_{\alpha}/\partial\phi) A^{\alpha}$] has been eliminated by the 
usingMaxwell's equations. During 
the inflationary phase, for $\eta < -\eta_1$,
the role of the conductivity 
shall be neglected. In this case 
the evolution equation for 
the canonical normal modes of the magnetic field
can be derived from the curl of Eq. (\ref{s2}) 
with the use of Eq. (\ref{s1b}): 
\begin{equation}
\vec{b}'' - \nabla^2 \vec{b} - \bigl[ \frac{1}{2}\frac{f''}{f} - 
\frac{1}{4} \bigl(\frac{f'}{f}\bigr)^2\bigr] \vec{b} =0,
\label{b}
\end{equation}
where $\vec{b} = \sqrt{f} \vec{B}$.
For $\eta > - \eta_1$ the effect of the conductivity
is essential. Therefore, the correct 
equations obeyed by the magnetic field will be 
the generalization of the MHD equations whose 
derivation will be now outlined.

MHD equations represent an effective description 
of the plasma dynamics for large length scales 
(compared to the Debye radius) and 
short frequencies compared to the 
plasma frequency. MHD can be derived 
from the kinetic (Vlasov-Landau) equations 
and the MHD spectrum 
indeed reproduces the plasma spectrum 
up to the Alvf\'en frequency \cite{k}. 
MHD 
can be also derived \cite{k} by neglecting the 
displacement currents in Eq. (\ref{s2}):
\begin{equation}
f \vec{\nabla}\times \vec{B} = \vec{J} + f' \vec{E}.
\end{equation}
By now using the Ohm law together 
with the Bianchi identity
we get to 
\begin{equation}
\bigl( 1 +  \frac{f'}{sf}\bigr) \vec{B}' = 
\vec{\nabla}\times(\vec{v} \times \vec{B}) + 
\frac{1}{s} \nabla^2 \vec{B}
\label{mh}
\end{equation}
which is the generalization of MHD 
equations to the case of evolving gauge coupling.
The quantity $s = \sigma /f$ is {\em constant}. 
The reason for this 
statement is the following. The rescaled conductivity,
\begin{equation}
\sigma = \sigma_c a \equiv \frac{T}{\alpha_{\rm em}},
\end{equation}
where $\alpha_{\rm em} \sim f^{-1} $. Therefore $\sigma/f =s$ 
with these rescalings, is constant. Taking 
now the Fourier transform of the fields appearing 
in Eq. (\ref{mh}) the solution, for the 
Fourier modes, will be 
\begin{equation}
B_{i}(k,\eta) = B_{i}(k,\eta_1) e^{- \int \frac{k^2 f}{ s f + f'} d\eta}.
\end{equation}

Consider now, as an example,  
\begin{equation}
f(\phi) = \bigl( \frac{\phi - \phi_1}{M_{P}}\bigr)^2.
\label{f}
\end{equation}
For $\eta < - \eta_1$ the solution of the evolution 
equation of the magnetic fluctuations
is, from Eq. (\ref{b}),
\begin{equation}
b(k, \eta) = N \sqrt{k \eta} H^{(2)}_{\nu}(k\eta),\,\,\, 
N= \frac{\sqrt{k\pi}}{2} e^{- 
i\frac{\pi}{4}(1 + 2\nu)},
\label{norm}
\end{equation}
where $N$ has been chosen in such a way that $b(k,\eta)\rightarrow 
\sqrt{k/2} e^{- i k
\eta}$ for $\eta\rightarrow -\infty$. Using Eq. (\ref{f}) 
 $\nu = 5/2$.

For $\eta_1 < \eta < \eta_{r}$ the Universe is reheating. 
During this phase the conductivity is not yet dominant
and the fastest growing solution outside the horizon
is given, in the case of Eq. (\ref{f}), by 
\begin{equation}
b(\eta) \sim \sqrt{f} \int_{\eta_1}^{\eta_{\rm r}} \frac{d\eta}{f}, 
\end{equation}
where we assumed, for concreteness, that $\alpha =2 $ in Eq. (\ref{rh}).
For $\eta > \eta_{\rm r}$ Eqs. (\ref{mh}) should be used.

The typical present frequency corresponding to the end of the inflationary 
phase is given, at the present time $\eta_0$, by 
\begin{equation}
\omega_{1}(\eta_0) \sim 10^{-4}\,\,T_{\rm dec}\,\,  
\epsilon^{\frac{1}{\alpha + 1}}\, 
\zeta^{\frac{\alpha -1}{2(\alpha + 1)}} \xi^{1/3}\,
\varphi^{- 2/3},
\label{om1}
\end{equation}
Notice that $10^{-4}\,T_{\rm dec}=100$ GHz where $T{\rm dec}$ 
is the decoupling 
temperature.  
The typical frequency corresponding to the 
onset of the radiation dominated phase is given by
\begin{equation}
\frac{\omega_{\rm r}(\eta_0)}{\omega_1(\eta_0)} \sim 
\biggl(\frac{\zeta}{\epsilon}\biggr)^{ \frac{1}{\alpha + 1}}.
\end{equation}

For $\eta > \eta_{\rm r}$ the conductivity 
dominates the evolution and using Eq. (\ref{mh}) we can estimate
the trace of the two-point function (\ref{two}) 
\begin{equation}
\rho_{B}(r,\eta) = \int \rho_{B}(k, \eta) \frac{ \sin{kr}}{k r} \frac{ dk}{k},
\end{equation}
with
\begin{equation}
\rho_{B}(k,\eta) =\frac{k^3}{\pi^2} |b(k,\eta)|^2. 
\end{equation}
Thus, in terms of
\begin{equation}
r(\omega) = \frac{\rho_{B}(\omega,\eta)}{\rho_{\gamma},
(\eta)}
\end{equation}
and using Eqs. (\ref{norm})--(\ref{mh})
\begin{equation}
r_{B}(\omega_{\rm G},\eta_0) = C(\nu,\omega_1,\omega_{\rm r})
\bigl(\frac{\omega_{\rm G}}{\omega_1}\bigr)^{5 - 2 \nu}
 {\cal T}(\omega_{\rm G})
\end{equation}
where $\omega_{\rm G} \sim 10^{-14} $Hz is the 
present frequency corresponding to a Mpc scale and
\begin{eqnarray}
&&C(\nu,\omega_1,\omega_{\rm r}) = \zeta^2
 \frac{2^{2 \nu - 2}}{\pi^2} \Gamma^2(\nu) 
\biggl(\frac{\omega_1}{\omega_{\rm r}}\biggr)^{4(\alpha +1)}.
\nonumber\\
&& {\cal T}(\omega_{\rm G}) = e^{ - \frac{\omega^2_{\rm G}}{\omega^2_{\sigma}}}
\bigl[ \bigl(\frac{\omega_{\vf}}{\omega_1} \bigr) 
\bigl( \frac{\omega_{\vf}}{\omega_{\rm m}}\bigr)^{\frac{1}{2}} 
\bigl( \frac{\omega_{\vf}}{\omega_{\rm c}}\bigr)^{\frac{3}{2}} \bigr]^{ 
2 \frac{\omega^2_{\rm G}}{T^2_0}},  
\label{aux}
\end{eqnarray}
with $\omega_{\sigma} \sim \sqrt{s/\eta_0}$ and $T_0 \sim 10^{-13}$ GeV.
In Eq. (\ref{aux}) $\omega_{\phi}$, $\omega_{\rm m}$ and $\omega_{\rm c}$ 
are, respectively, the present values of $H_{\phi}$, $H_{\rm m}$ and 
$H_{\rm c}$. Using the notation of Eq. (\ref{om1}) we have that
\begin{eqnarray} 
&&\omega_{\rm m} \sim \epsilon^{-1/(\alpha + 1)}\, \xi^{1/2}\, \zeta^{(1- 
\alpha)/(2 \alpha + 2)}\, \omega_1
\nonumber\\
&&\omega_{\rm c}  \sim \varphi^2 
\epsilon^{-1/(\alpha + 1)}\, \xi^{1/2}\, \zeta^{(1- 
\alpha)/(2 \alpha + 2)}\, \omega_1
\nonumber\\
&&\omega_{\phi} \sim 
\xi^{7/6} \varphi^{2/3}  \epsilon^{-1/(\alpha + 1)}\, \xi^{1/2}\, \zeta^{(1- 
\alpha)/(2 \alpha + 2)}\, \omega_1.
\end{eqnarray} 
All the frequencies are evaluated at the time $\eta_0$.
Using now the previous equations,
\begin{equation}
r_{B}(\omega_{\rm G}) \sim \epsilon^4 \zeta^{-2}.
\label{rb}
\end{equation}
This result should be confronted with typical values of $r_{\rm B}$ required
in order to explain galactic (and possibly inter-galactic) magnetic fields.

Large scale magnetic fields are a well known 
component of the interstellar medium. 
The most reliable 
estimates available today rely on Faraday rotation 
of radio signals. With these techniques the magnetic fields 
of different galaxies have been measured both within 
and beyond our local group \cite{kro}. Recently magnetic fields 
in clusters have been shown to possess a magnetic field 
which is larger than previously thought and of the 
order of the $\mu$ Gauss \cite{cl}. These measurements 
have been made possible through the 
combined analysis of x-rays bright Abell clusters
performed with VLA telescope and ROSAT satellite 
full sky survey. 

Through 
differential rotation of the primeval galaxy some small 
magnetic seeds can be substantially amplified. Since by the time 
of the formation of the galaxy the gauge coupling is frozen 
the correct equation describing large scale magnetic will be exactly Eq. 
(\ref{mh}) with $f'=0$ and $s\rightarrow \sigma$. In this equation 
an instability develops when the dynamo term (containing the
bulk velocity) dominates over the diffusivity term (containing 
the conductivity). 
 
Taking into account that the rotation 
period is of the order of $10^{8}$ yrs and that 
the age of the galaxy is of the order of 
$10^{10}$ yrs the typical amplification 
which could be obtained through the dynamo mechanism 
is of the order of 30 e-folds, namely, $13$ orders of 
magnitude. This implies that if today galactic 
magnetic fields have $\mu$ Gauss strength, at the end of
 gravitational collapse they should have been as small 
as $10^{-19}$ Gauss in order to turn on the dynamo mechanism.

When the primeval galaxy collapses 
(from a typical scale of $1$ Mpc down to a scale of $30$ kpc)
the frozen-in magnetic flux increases of roughly $4$ orders of 
magnitude. This is because the mean density 
prior to collapse is of the order of the critical 
density, whereas, after collapse, the mean density 
of the galaxy is approximately six orders of magnitude larger.
These estimates have been obtained for $h=0.65$, $\Omega_{\Lambda} =0.7$ 
and $\Omega_{\rm matter} = 0.3$. 

Taking now into account this last point the primordial seeds should 
be as strong as $10^{-23}$ Gauss over a typical scale $\omega_{\rm G}^{-1}$ 
prior to gravitational collapse. In terms of $r_{B}(\omega)$
this means that in order to turn on the dynamo mechanism 
we should have \cite{tw}
\begin{equation}
r_{B}(\omega_{\rm G} ) \geq 10^{-34}.
\label{G}
\end{equation}
Various mechanisms have been proposed so far in order to explain 
the magnetic field of the galaxy \cite{M}. 
Recently \cite{mg}, it was pointed out that the evolution of gauge couplings 
in a Kaluza-Klein context can offer interesting possibilities for  
the generation of primordial magnetic fields. The present analysis 
could be viewed as a realization of that proposal in a geometry 
compatible with a de Sitter stage of inflation.

In the case of clusters the dynamo mechanism 
is more problematic. Indeed on one hand clusters rotate less 
than galaxies and, on the other hand, the mean density of matter 
is smaller. Therefore, larger values of $r_{B}$ 
are required in order to successfully implement the dynamo 
mechanism for clusters.

\begin{figure}
 \centerline{\epsfxsize = 8 cm  \epsffile{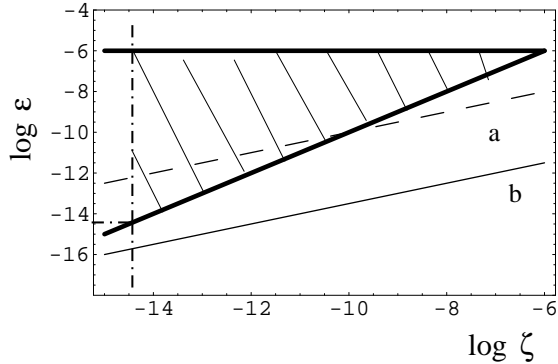}} 
\caption[a]{Magnetogenesis requirements 
are illustrated. The thin line {\bf b} corresponds to 
$r_{\rm B}(\omega_{\rm G}) \sim 10^{-34}$. The dashed line {\bf a} 
corresponds $r_{\rm B}(\omega_{\rm G}) \sim 10^{-20}$. The (horizontal)
 thick line correspond to $\epsilon = 10^{-6}$. The (diagonal) thick line 
corresponds to $\epsilon \sim \zeta$. Magnetogenesis is possible when the parameters 
lie in the shaded area.}
\label{F1}
\end{figure}

Using Eqs. (\ref{rb}) and (\ref{G}) the 
allowed region in the parameter space of the model can be obtained by taking 
into account the constraints discussed in the previous section.
In Fig. \ref{F1} the shaded area illustrates the region where 
magnetogenesis is possible.
To be consistent with inflationary production
of scalar and tensor fluctuations of the geometry $\epsilon < 10^{-6}$ 
should be imposed. Thus, in Fig. \ref{F1} the parameters 
should all lie below the (horizontal) thick line. Moreover, since 
$H_{\rm r} < H_{1}$, $\zeta < \epsilon$. Recall that $\zeta > 10^{-15}$  
in order not to affect the nucleosynthesis epoch [see the dot-dashed line 
in Fig. \ref{F1}]. This requirement comes about since $H_{\rm r} > m$ and
$\xi \geq 10^{-15}$.

In this paper the possible phenomenological implications of the 
evolution of the gauge coupling have been analyzed in the case 
of a specific model of background evolution. It has been shown that
if the gauge coupling is related to a massive scalar field minimally coupled 
to the geometry the phenomenological constraints related both to the 
evolution of the massive scalar and to the evolution of the gauge coupling 
can be satisfied. In a specific example the large scale magnetic fields 
produced with this mechanism have been computed. It has been shown that 
they can be large enough to seed the galactic dynamo mechanism. They can be 
also relevant for the origin of magnetic fields in clusters. Further 
work on these possibilities is in progress.

{\em Acknowledgments} The author wishes to thank M. E. Shaposhnikov 
for important discussions.

\end{document}